# MIoT-Driven Comparison of Open Blockchain Platforms

Abdou-Essamad Jabri [1, 0009-0003-3510-0087], Mostafa AZIZI [1, 0000-0001-9936-3001], Cyril DROCOURT [2, 0000-0003-1636-9462], and Gil UTARD [2, 0000-0003-3081-6185]

[1] MATSI, ESTO, UMP, Oujda, Morocco
[2] MIS, UPJV, Amiens, France
{abdou-essamad.jabri, azizi.mos}@ump.ac.ma
cyril.drocourt@u-picardie.fr
gil.utard@u-picardie.fr

**Abstract.** Being propelled by the fourth industrial revolution (Industry 4.0), IoT devices and solutions are well adopted everywhere, ranging from home applications to industrial use, crossing through transportation, healthcare, energy, and so on. This wide use of IoT has not gone unnoticed, hackers are tracking the weakness of such a technology and threatening them continuously. Their security at various levels has become an important concern of professionals and researchers. This issue takes more risk, especially with the IoT variants, IIoT (Industrial IoT) and MIoT (Medical IoT). Many existing security solutions are adapted and proposed for addressing IoT security. In this paper, we are interested in exploring blockchain technology and we make a comparison of three free Blockchain platforms towards their applicability for MIoT context, namely Ethereum, Hyperledger Fabric and Corda. In general, Blockchain technology provides a decentralized, autonomous, trustless, and distributed environment. It is challenging to find a Blockchain platform that fits the MIoT context and performs well in terms of security. The retained platform should be deployed smartly to avoid its practical drawbacks related to energy-consuming and excessive computing.

**Keywords:** IoT, MIoT, Blockchain, Security, Ethereum, Ledger Fabric, Corda.

## 1. Introduction

During the last decades, many technologies have emerged such as IoT (Internet of things), Big Data, Modern AI, and so on. IoT with its involvement in data communication, is gaining ground on all horizons. Its presence in many domains, such as IIoT and MIoT, relies on the use of devices with sensors to collect data and share it through public or private networks with low energy consumption [1]. Despite all these qualities, IoT remains still vulnerable to a wide range of attacks due to its limited resources (e.g. processor, storage, and network capacity) [2] to address all those security challenges; to this end, the blockchain is foreseen as a potential solution. As known, the blockchain is a chaining of blocks; each block contains structured data of transactions between peers interconnected in a blockchain network (like Ether in Ethereum). All these blocks are hashed using cryptographic functions to guarantee their confidentiality and integrity, and are recorded, as proof of all the witnessed transactions, in an immutable distributed ledger.

This technology was first used to allow direct transactions between parties without going through a third party (such a financial institution), and to prevent double-spending in a public peer-to-peer network that anyone can get in as long as they accept the record of transactions before joining [3]; this led to the creation of the most known cryptocurrency: Bitcoin. Next, new blockchains have been created with more flexibility and diversity. One of the most popular is Ethereum that can create more relevant transactions due to its capability to store, not only assets of the transactions, but also a set of codes called smart contracts. Written in the programming language Solidity, the Ethereum smart contracts are used to automate the execution of agreements or transactions, enabling trustless and secure interactions between involved parties. Generally, the birth of smart contracts helped so much the developers by allowing more flexibility in controlling, not only transactions, but even the creation of private or permissioned networks with well-defined smart contracts, implementing access policy to specific resources. This new approach opens the door for the Blockchain technology to be implemented in other domains, such as securing IoT infrastructures dedicated for healthcare (see Figure 1). In this paper, we present a comparison between three of the most popular permissioned blockchain networks, namely Ethereum, Hyperledger Fabric and Corda; this task is conducted by the vision of their usefulness for securing MIoT infrastructures. To this end, we use some criteria of performance regarding time, energy consumption, friendliness, and smart contract coding.

The rest of this paper is organized as follows: Section 2 delves into the literature review of related works over mostly using blockchains and securing IoT infrastructures. While Section 3 deals with the comparison of three selected open blockchain solutions, using different criteria. Section 4 concludes this work by stating some recommendations and standing some perspectives.

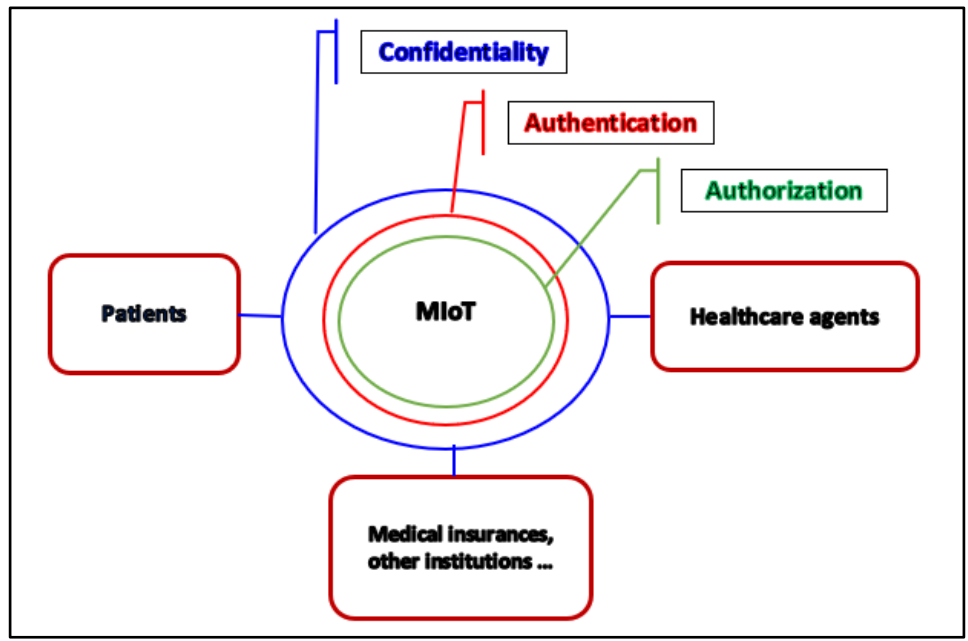

Figure 1. Our simplified vision of securing MIoT infrastructures

## 2. Related works

### 2.1. Selection of papers

This section describes our methodology for reviewing recent papers on IoT and Blockchain in the medical context. Preferred Reporting Items for Systematic Reviews and Meta-Analysis (PRISMA) guidance were followed. The search for articles was conducted on the Web of Science and ScienceDirect. Therefore, to select the papers considered for this review, we have applied this search criteria "IoT AND Security AND Blockchain," "(Medical IoT OR healthcare) AND Blockchain," and "Ledger Fabric OR Ethereum OR Corda" over the last five years. At the end, we got 10 papers. All these works [4-12] will be examined in the following subsection.

### 2.2. Review of the selected papers

### A- Targeting MIoT Security

The entry of IoT into the medical domain has made a massive impact in terms of remote monitoring using fixed and wearables devices that give a real-time observation and surveillance of the Physiological parameters which really help the detection and quickly respond to any health problem that may threaten the patient's life [4]. This impact was seen clearly in the covid-19 pandemic where IoT played a significant role under the umbrella of contact tracking applications to draw the spread of the virus map and quarantine affected areas [5]. Those devices can also reduce time and cost in hospital's infrastructures by controlling temperature, improving surveillance which will reduce staff cost and hospital admissions [6]. On the other hand, IoT devices necessitate specific requirements that focus on security, privacy, and compatibility; Butpheng et al. [7] gave an overview on security and privacy risk that came with IoT devices through different parts of the architecture. Obaid et al. [8] point out the importance of securing the process of collecting and sharing the information of patients. Affia et al. [9] highlight the drawbacks of those devices with unencrypted communication or weak authentication that may lead to a denial of healthcare services.

B- **Targeting the three blockchain platforms**

To address those security concerns, many researchers suggest using blockchain technology. This approach comes with a distributed immutable ledger capable of reinforcing the security posture of IoT and its variants IIoT [10] and MIoT [11]. However, the diversity of blockchain solutions necessitates comparing them. Pancari et al. [12] compare Ethereum and Hyperledger fabric in terms of implementing Attribute-Based Access Control (ABAC) in the Smart home concept. Wilson et al. [13] analyze the capabilities of Ethereum and Hyperledger Fabric in terms of security, integrity, and scalability. Valenta et al [14] compare the three blockchain platforms citing each one consensus mechanisms, its pros, and cons without forgetting their capability to secure IoT devices. Based on this first look we easily recognize that Hyperledger fabric and Ethereum had been more usable than Corda in terms of securing IoT devices.

## 3. Comparison of Ethereum, Hyperledger Fabric and Corda

### 3.1. Procedure of comparison

First, we define the requirements of medical IoT devices using recent papers, focusing on security, privacy, and compatibility to get the relevant criteria for our comparison. Next, we gather information on each platform using official documentation and research papers that present the blockchain as a security solution for MIoT. This gives us a better understanding of each platform, namely the pros and cons regarding our target. The collected information about these three platforms is gathered within a table (see Table 1) and analyzed based on the targeted criteria:
- Compatibility and support of our use case: define which platform is more compatible with medical IoT solutions.
- Performance: which refers to time of implementation and execution, Energy consumption and smart contracts configuration.
- Flexibility: Refers to the degree of how much the platform can be adapted to project's need

### 3.2. Results of comparison

Table 1. Aspects of comparison between Ethereum, Hyperledger Fabric, and Corda

| Aspect of Comparison | Ethereum | Hyperledger Fabric | R3 Corda |
|---|---|---|---|
| **Support for Medical IoT Devices** | Easier integration with existing tools and frameworks | limited in the pre-build, but its flexible architecture allows for integration with various IoT protocols using specific tools or plugins. | Primarily targeted at financial applications and has limited direct support for Medical IoT devices. However, it can be customized through additional integrations. |
| **Time of implementation** | the implementation might take time due to the lack of pre-configured infrastructure | faster implementation due to pre-configured infrastructure that can be customized. | faster implementation due to pre-built templates and tools specifically designed for setting up private networks. |
| **Cost of implementing smart contracts** | Gas generation and consumption for transactions and smart contract execution, resulting in higher system demands | No gas consumption, leading to lower system strain and easier interaction with network | Network fees might be implemented, but typically lower than gas fees. |
| **Coding of Smart contracts** | Specific programming language: *Solidity* | Standard programming language: *Java* and Go | Standard programming language: *Java* and *Kotlin* |
| **Energy consumption** | High due to PoW but with the new consensus base on PoS the system is more energy-efficient | PBFT require less energy then Pow but when it comes to PoS the specific implementation and the network size plays a major role | the most energy-efficient between the cited platforms |
| **Execution Time** | Slower due to the need of confirmation of each block | faster due to the pre-selected validators | faster than other due to the reliability on trusted relationships between participants |
| **Flexibility** | pre-defined standards and the public nature make it limited | high flexibility due to its modular architecture | flexible in specific cases |

## 3.3. Discussion

Ethereum was using the Proof of Work (PoW) [15] as a consensus mechanism, where miners compete to solve complex mathematical puzzles to validate transactions and create new blocks. This process is resource-intensive and can be slow, as each block takes around 10 minutes to be added to the blockchain. However, Ethereum has switched to a Proof of Stake (PoS) [15] as a consensus mechanism, which requires validators to lock up a certain amount of Ether as collateral to validate transactions and create new blocks. This remarkable change has improved energy efficiency and the execution time. Hyperledger Fabric: Hyperledger Fabric uses a modular consensus mechanism that allows organizations to choose between different consensus algorithms based on their specific needs. The default consensus algorithm is Practical Byzantine Fault Tolerance (PBFT) [17], which is based on a voting process among a subset of nodes. This system is built for enterprise use cases and provides high throughput and low latency for transaction processing. Corda [18] uses a unique consensus mechanism called the Uniqueness Consensus, where only the parties involved in a transaction need to reach consensus. This allows for private and secure transactions between parties without the need for a global consensus mechanism which results in less execution time and energy consumption, Corda also allows for the creation of custom consensus mechanisms based on the specific needs of different use cases.
 In Ethereum, smart contracts are written in the programming language Solidity (see Figure 2). These contracts are only deployed on the Ethereum blockchain and are executed by all nodes in the network when triggered by a transaction. Smart contracts on Ethereum are used to automate the execution of agreements or transactions, enabling trustless and secure interactions between parties. In Hyperledger Fabric, smart contracts are referred to as "chaincode" (see Figure 3). Chaincode in Hyperledger Fabric [17] can be written in various programming languages such as Go, Java, or JavaScript. Chaincode runs on specific channels within the Hyperledger Fabric network and is only invoked by the parties involved in a particular transaction. This allows for more privacy and scalability compared to Ethereum. In Corda, smart contracts (see Figure 4) are known as "CorDapps" [18]. These are written in Java or Kotlin and are only shared between the parties directly involved in the transaction. CorDapps in Corda are designed to enforce the terms of the agreement between parties and facilitate the exchange of assets or information. Corda's architecture allows for more privacy and scalability compared to Ethereum and Hyperledger Fabric.

We have presented in the three figures (see Figures 2, 3, and 4) the example about reading temperature from an IoT device; to transfer it, it is submitted as transaction to the blockchain ledger to be recorded, and it is accessible from there by retrieval.

```solidity
...
contract IoTContract {
    struct TemperatureRead {
        string deviceId;
        uint256 timestamp;
        uint256 temperature;
    }

    mapping(string => TemperatureRead) private temperatureRead;

    function submitTemperatureRead(string memory deviceID, uint256 temperature) external {
        TemperatureRead memory read = TemperatureRead({
            deviceId: deviceID,
            timestamp: block.timestamp,
            temperature: temperature
        });

        temperatureReads[deviceID] = read;
    }

    function getLatestTemperatureRead(string memory deviceID) external view returns (TemperatureReading memory) {
        return temperatureReads[deviceID];
    }
}
```

**Figure 2. Example of an IoT contract for managing IoT temperature data on Ethereum with Solidity**

```go
...
type MIoTContract struct {
    contractapi.Contract
}
type TemperatureRead struct {
    DeviceID    string  `json:"deviceId"`
    Timestamp   int64   `json:"timestamp"`
    Temperature float64 `json:"temperature"`
}
func (c *IoTContract) SubmitTemperatureRead(ctx contractapi.TransactionContextInterface, deviceID string, temperature float64) error {
    // Get the current timestamp
    timestamp, err := ctx.GetStub().GetTxTimestamp()
    ...
    // Create a new TemperatureRead
    read := TemperatureRead{
        DeviceID:    deviceID,
        Timestamp:   timestamp.Seconds,
        Temperature: temperature,
    }
    // Convert the TemperatureReading to JSON
    readJSON, err := json.Marshal(read)
    ...
    // Save the JSON to the ledger
    ctx.GetStub().PutState(deviceID, readJSON)
    ...
}
func (c *IoTContract) GetLatestTemperatureRead(ctx contractapi.TransactionContextInterface, deviceID string) (*TemperatureRead, error) {
    // Retrieve the state from the ledger
    readJSON, err := ctx.GetStub().GetState(deviceID)
    ...
    }
...
return &read, nil
}
func main() {
    // Create a new instance of the contract
    contract := new(IoTContract)
    // Create a new Chaincode
    cc, err := contractapi.NewChaincode(contract)
    ...
    // Start the Chaincode
    cc.Start();
    ...
}}
```

**Figure 3. Example of an IoT contract for managing IoT temperature data on Hyperledger Fabric with Go**

```kotlin
package com.template

import net.corda.core.contracts.Command
import net.corda.core.contracts.CommandData
import net.corda.core.contracts.Contract
import net.corda.core.contracts.requireSingleCommand
import net.corda.core.transactions.LedgerTransaction

data class TemperatureRead(val deviceId: String, val timestamp: Long, val temperature: Double)

class IoTContract : Contract {
    companion object {
        @JvmStatic
        val IO_CONTRACT_ID = "com.template.IoTContract"
    }

    override fun verify(tx: LedgerTransaction) {
        val command = tx.commands.requireSingleCommand<Commands>()

        when (command.value) {
            is Commands.SubmitTemperatureRead -> {
                require(tx.inputStates.isEmpty()) { "No input states should be consumed when submitting a temperature Read." }
                require(tx.outputStates.size == 1) { "Exactly one output state should be created when submitting a temperature Read." }

                val output = tx.outputStates.single() as TemperatureRead
                require(output.deviceId.isNotEmpty()) { "Device ID must not be empty." }
                require(output.temperature >= 0) { "Temperature must be a non-negative value." }
            }
            else -> throw IllegalArgumentException("Unrecognized command: ${command.value}")
        }
    }

    interface Commands : CommandData {
        class SubmitTemperatureRead : Commands
    }
}
```

**Figure 4. Example of an IoT contract for managing IoT temperature data on Corda with Kotlin**

## 4. Conclusion

In this paper, we have discussed the issue within the IoT infrastructures in general, and particularly those deployed in the medical context (MioT). Indeed, several actors are involved over these platforms in different use cases. These actors include patients, doctors, nurses, and insurance companies. The data streaming between these parties should respect security requirements, covering privacy and data protection. In this sense, we have surveyed three blockchain platforms, namely Ethereum, Hyperledger Fabric, and Corda. As a result of the comparison between them, we find out that Hyperledger Fabric is the fittest one to our issue with MIoT.